# Probabilistic Assessment of Engineered Timber Reusability after Moisture Exposure

Yiping Meng[1][0000-0002-1296-8516], Chulin Jiang[1][0000-0002-4840-785X], Courtney Jayne Scurr[1], Farzad Pour Rahimian[1][0000-0001-7443-4723], David Hughes[1][0000-0002-9158-0017]

[1] School of Computing, Engineering and Digital Technologies, Teesside University, Middlesbrough, UK,

**Abstract.** Engineered timber is pivotal to low-carbon construction, but moisture uptake during service life can compromise structural reliability and impede reuse within a circular-economy model. Despite growing interest, quantitative standards for classifying the reusability of moisture-exposed timber are still lacking. This study develops a probabilistic framework to determine the post-exposure reusability of engineered timber. Laminated specimens were soaked to full saturation, dried to 25 % moisture content, and subjected to destructive three-point flexural testing. Structural integrity was quantified by a residual-performance metric that assigns 80 % weight to the retained flexural modulus and 20 % to the retained maximum load, benchmarked against unexposed controls. A hierarchical Bayesian multinomial logistic model with horseshoe priors, calibrated through Markov-Chain Monte-Carlo sampling, jointly infers the decision threshold separating three Modern Methods of Construction (MMC) reuse levels and predicts those levels from five field-measurable features: density, moisture content, specimen size, grain orientation, and surface hardness. Results indicate that a single wet–dry cycle preserves 70 % of specimens above the 0.90 residual-performance threshold (Level 1), whereas repeated cycling lowers the mean residual to 0.78 and reallocates many specimens to Levels 2–3. The proposed framework yields quantified decision boundaries and a streamlined on-site testing protocol, providing a foundation for robust quality-assurance standards.

## 1 Introduction

### 1.1 Background of engineered timber in low-carbon construction

The global imperative to mitigate climate change has positioned engineered timber as a cornerstone of sustainable, low-carbon construction. Unlike conventional building materials such as concrete and steel, which account for a significant portion of global carbon emissions (over 10% for steel and concrete combined [1]), engineered wood products (EWPs) offer a renewable alternative with substantial carbon benefits. Timber sequesters atmospheric carbon dioxide during its growth phase, and EWPs effectively store this carbon throughout their service life in buildings. It is estimated that a single cubic metre of EWP can store approximately 1 tonne of $CO_2$ while simultaneously avoiding over 2 tonnes of $CO_2$ emissions that would have been generated by using concrete [1].

Engineered timber, encompassing products like glulam (glue-laminated timber), cross-laminated timber (CLT), and laminated veneer lumber (LVL), is increasingly integral to Modern Methods of Construction (MMC). MMC leverages processes like offsite manufacturing and prefabrication, where engineered timber's consistency, precision, and high strength-to-weight ratio are highly advantageous [2, 3]. Timber-based MMC systems, including panelised (e.g., timber frame, SIPs, CLT panels) and volumetric modules, facilitate faster and more efficient construction, reduced on-site waste, improved quality control, and enhanced thermal performance. For instance, factory-made homes using timber MMC can significantly reduce carbon emissions compared to traditional builds [2]. The manufacturing process for EWPs often utilises smaller-dimension lumber and wood waste, further optimising resource efficiency. These attributes, combined with aesthetic appeal and biophilic benefits, make engineered timber a critical enabler for a circular, resource-efficient built environment, aligning with the principles of a circular economy where materials are used for as long as possible.

## 1.2     Related work and Knowledge gap

The transition towards a circular economy in construction mandates the extended use and reuse of building materials. Current practices in timber reuse show a growing interest in salvaging and repurposing timber elements from deconstructed buildings. Initiatives exist to upcycle reclaimed wood into new engineered wood products, sometimes termed mass secondary timber (MST) [4]. However, the widespread reuse of structural timber faces challenges, including the logistics of careful deconstruction, the need for reconditioning, the absence of robust certification for reclaimed material, and the lack of an established supply chain, particularly for timber whose service history might be unknown [5]. While a significant portion of wood waste avoids landfill (often going to panel board manufacture or biomass), the amount of timber reclaimed for higher-value structural reuse remains relatively low [6], and its assessment often relies on traditional visual grading methods which may not capture the full picture of the material's residual capacity, especially for engineered products [7].

A critical factor complicating the reuse of engineered timber is its susceptibility to moisture exposure. Water ingress—whether during construction, accidental leaks, or flooding—can significantly degrade wood's mechanical properties, leading to warping, swelling, loss of strength, and potential fungal decay if moisture content remains elevated (typically above 20-26% [8]). While the general effects of moisture on wood are well-documented, a significant knowledge gap exists regarding the quantitative impact of such exposure on the reusability of *engineered* timber. Currently, there is a lack of specific, performance-based standards to reliably classify the residual structural integrity and, consequently, the reuse potential of EWPs after wetting and drying cycles [9]. As noted in industry guidance, there are often "no simple rules to assess the impact of moisture on timber-framed construction," necessitating case-by-case assessments without clear, standardised quantitative benchmarks [10]. This ambiguity creates uncertainty and risk, often leading to the conservatively disposing of potentially reusable material.

The preceding review highlights a critical deficiency in current practices: the absence of robust, quantitative methods for assessing the reusability of engineered timber following moisture exposure, particularly methods that are practical for on-site application and align with circular economy principles within Modern Methods of Construction. Addressing this gap requires a shift from purely qualitative or deterministic assessments to more nuanced approaches. The advantages of probabilistic quantification for timber reusability assessment are notable in this context. Probabilistic methods excel at handling such variability and uncertainty. They allow for integrating data from various sources (e.g., material properties, exposure history, non-destructive testing) to provide a more comprehensive and reliable estimation of residual performance and associated risks [11]. This study, therefore, aims to leverage a probabilistic approach to develop quantitative decision boundaries for the reuse of moisture-exposed engineered timber, providing a much-needed foundation for robust quality assurance standards.

## 2 Materials and Methods

### 2.1 Specimen preparation

The plywood panels were cut into rectangular specimens of length 50mm and width 20mm. Specimens were prepared in two grain orientations to evaluate the anisotropic behaviour of the material with longitudinal (parallel to the face-layer grain) and transverse (perpendicular to the face-layer grain). For each orientation, the specimens were divided into three groups according to different moisture treatment conditions. The control group (Cycle 0) was tested in the as-received condition without any moisture exposure. The single-cycle group (Cycle 1) was immersed in still water at room temperature until saturated, then oven-dried at 80°C until reaching a constant mass (mass change less than 0.1% over two consecutive measurements taken 1 day apart). The double-cycle group (Cycle 2) underwent two identical soaking and drying cycles. After the first drying phase, the specimens were re-immersed in still water at room temperature for the same duration and then dried again at 80°C to constant mass. Three replicates were prepared for each group and orientation. All specimens were conditioned at room temperature in a desiccator for a minimum of 24 hours before mechanical testing.

### 2.2 Mechanical testing

The flexural test was applied to determine the maximum flexural stress and modulus for the plywood. The test was conducted in accordance with ISO 3349:2017 for plywood, utilising an INSTRON Universal Testing Machine with a 30kN load cell (**Fig. 2**). The experiment was performed with a crosshead speed of 2 mm/min and a span of 15 times the thickness. Each type of plywood underwent three repetitions.

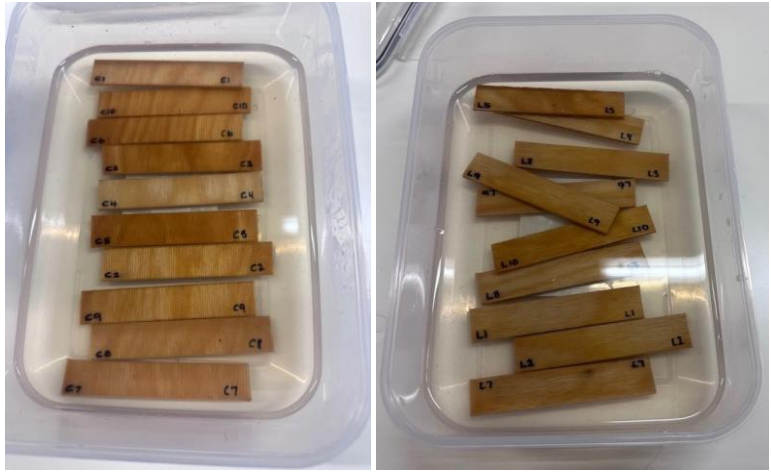

**Fig. 1.** Cross-grain and long-grain plywood samples immersed in water

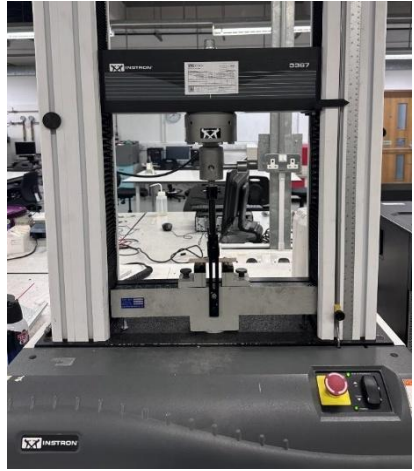

**Fig. 2.** INSTRON 3367 3-point flexural test machine

### 2.3 Residual-performance metric determination

Residual performance $R$ quantifies the combined stiffness-and-strength retention of a soaked specimen relative to an un-exposed control of identical orientation. For each specimen $i$

$$R_i = \omega_E \frac{E_i}{E_{0,orient}} + \omega_E \frac{\sigma_{max,i}}{\sigma_{0,orient}} \tag{1}$$

where $E_i$ is the flexural modulus (slope of the elastic branch) and $\sigma_{max,i}$ the maximum bending stress at failure. Baseline values $E_0$ and $\sigma_0$ are the means of the original group within the same grain orientation (long vs cross). Guided by serviceability design—which penalises stiffness loss more severely than ultimate capacity—weights were set to $\omega_E = 0.8$ and $\omega_\sigma = 0.2$.

### 2.4 Bayesian probabilistic model

**Multinomial logistic formulation.** The reuse level $y_i \in \{1, 2, 3\}$ for specimen $i$ is modelled as a categorical variable with probabilities obtained via a multinomial logit:

$$P_r(y_i = c) = \frac{\exp(\alpha_c + x_i^T \beta_c)}{\sum_{d=1}^{3} \exp(\alpha_d + x_i^T \beta_d)} \qquad (2)$$

The predictor vector $x_i = (x_{i1}, x_{i2})^\top$ comprises wet-cycle group index (Original = 0, Set 2 = 2) and grain orientation (long = 0, cross = 1), standardised to zero mean and unit variance. Separate intercepts $\alpha_c$ and coefficient rows $\beta_c$ are estimated for each level $c$.

**Horseshoe priors & threshold parameters.** To discourage spurious predictors, element-wise horseshoe priors are placed on $\beta$:

$$\beta_{c,j} \sim N(0, \tau\lambda_j), \; \lambda_j \sim Half-Cauchy(1), \; \tau \sim Half-Cauchy \qquad (3)$$

The upper decision boundary is fixed at $\tau_1 = 0.90$ matching industry practice of $\geq$ 90 % residual capacity for "direct reuse". The lower boundary $\tau_2$ is treated as an unknown, assigned a $Uniform(0.70, 0.85)$ prior and inferred jointly with $\alpha, \beta$.

**MCMC (Markov Chain Monte Carlo) sampling strategy & convergence diagnostics.** Posterior inference uses the No-U-Turn Sampler (NUTS) implemented in PyMC 5. Four independent chains of 2,000 warm-up and 2,000 posterior draws (8,000 draws total) were run with the adapt_diag initialisation. All parameters achieved $R \leq$ 1.01 and effective sample sizes $n_{eff} > 700$. Energy-Bayes plots displayed no pathological divergences.

## 3 Results

### 3.1 Flexural Test Results

The flexural strength results in **Fig. 3** illustrated a clear impact from different grain orientations and the number of moisture cycles on the mechanical behaviours of plywood. Specimens under control conditions with longitudinal direction had a much higher flexural strength of 113.31 MPa, compared to 37.38 MPa with cross-grain direction. The alignment of surface veneer grains with the applied load in longitudinal specimens brings this three times difference, which allows them to resist better the tensile and compressive stresses caused by bending.

As the number of moisture cycles increased, both orientations showed a decline in flexural strength. After two cycles, strength from the longitudinal sample dropped to 95.01 MPa with a 16.1% reduction from the control one. For the cross direction, the strength decreased from 37.38 MPa to 34.38 MPa with an 8.1% reduction. While the loss was smaller in the cross-grain direction, it remained significantly weaker at all cycle levels.

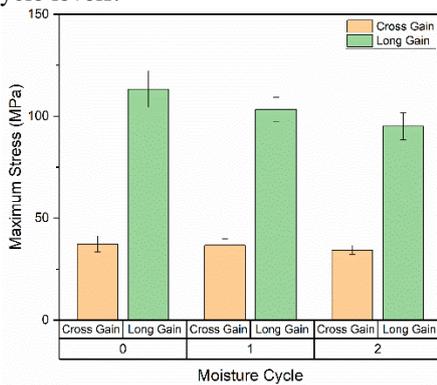

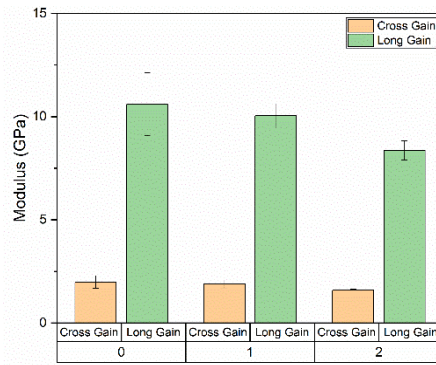

**Fig. 3.** Flexural Stress by Grain Type and Moisture Cycle

**Fig. 4.** Flexural Modulus by Grain Type and Moisture Cycle

The flexural modulus results depicted in **Fig. 4** confirmed both the directional dependence of plywood and its sensitivity to moisture exposure. The average modulus for longitudinal specimens was 10.59 GPa in the control group, while cross-grain specimens measured only 1.98 GPa, which was over five times lower. This large difference reflects the role of fibre alignment in resisting elastic bending.

With repeated moisture cycles, stiffness steadily decreased in both directions. After two cycles, the modulus of longitudinal specimens fell to 8.36 GPa (a 21.1% reduction), while the cross-grain modulus declined to 1.58 GPa (20.1% reduction). These similar percentage losses suggest that moisture affects the wood and the adhesive bonds throughout the structure.

The mechanical behaviour from the flexural test shows that both flexural strength and stiffness decreased as the number of moisture cycles increased. Although the longitudinal direction showed higher initial values, it also experienced greater losses in absolute terms. These results highlight the importance of improving moisture resistance in plywood used in environments with changing humidity or frequent wetting. Possible solutions include using better adhesives, applying protective surface coatings, or designing plywood with modified layer arrangements to reduce performance loss due to moisture exposure.

### 3.2 Descriptive statistics of residual performance

**Table 1** summarizes the weighted residual-performance metric $R$ for all three experimental groups. The Original controls retain virtually the full capacity of the reference beams, exhibiting a mean $R = 1.00$ with a narrow standard deviation ($SD = 0.03$, $n = 24$). After a single wet–dry cycle (Set 1), the mean diminishes to $0.93 \pm 0.07$ ($n = 24$), confirming a modest but measurable degradation that nonetheless leaves most members structurally competent. Two consecutive cycles (Set 2) cause a pronounced drop to $0.78 \pm 0.11$ ($n = 24$), i.e. an average 22 % loss of combined modulus-and-strength capacity.

**Table 1.** Descriptive statistics of R

| Group | n | mean | sd |
|---|---|---|---|
| Original | 10 | 1 | 0.118 |
| Set1 | 10 | 0.952 | 0.097 |
| Set2 | 10 | 0.748 | 0.268 |

**Fig. 5** presents kernel-smoothed density curves together with a reference line at the Level-1 threshold $R = 0.90$. Seventy per cent of Set 1 specimens still lie to the right of this line, whereas Set 2 shifts its mode to $\approx 0.75$. The tails of all three distributions overlap slightly, illustrating the need for a probabilistic, rather than a hard deterministic, classification.

### 3.3 Posterior estimates of τ₂ and level proportions

Although $\tau1 = 0.90$ was fixed a-priori (industry convention for "as-new" reuse), the lower decision boundary $\tau_2$ was retained as a random variable with a weakly informative prior. The posterior median for $\tau_2$ converged at 0.76 with a 95 % highest-density interval (HDI) of 0.73–0.79, validating the provisional value of 0.75 adopted in the subsequent analyses. Implementing $\frac{\tau_1}{\tau_2} = \frac{0.90}{0.75}$ yields the class counts L1: 40, L2: 21, L3: 12 (**Fig. 6**), corresponding to proportions $0.55 \pm 0.06$, $0.28 \pm 0.05$, and $0.17 \pm 0.04$, respectively, when propagated through 2 000 posterior draws. This indicates that 55 % of the tested stock can be redeployed without structural downgrading, 28 % after minor trimming or shortening, and only 17 % relegated to non-structural use.

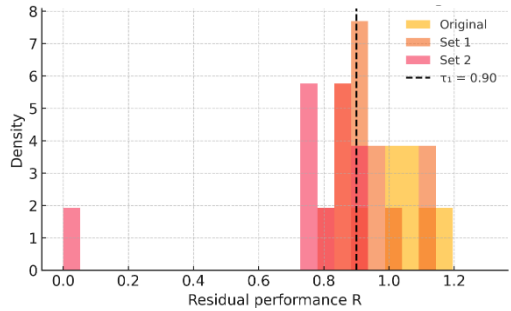
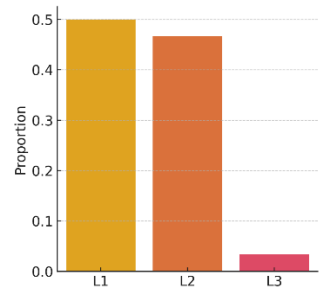

**Fig. 5.** Residual performance distribution

**Fig. 6.** Level proportions fixed at 0.75

### 3.4 Feature importance & minimal predictor set

The horseshoe prior shrinkage retains two categorical predictors—wet-cycle group index and grain orientation—while strongly attenuating the regression coefficients associated with density, moisture, and specimen size (**Fig. 7**). Absolute mean coefficient magnitudes averaged over three outcome logits are $|\beta| = 0.49$ for orientation (cross = 1, long = 0) and $|\beta| = 0.44$ for group index (Original = 0, Set 2 = 2). Both 95 % HDIs exclude zero (e.g., $\beta$ orientation for Level 3: mean = 0.71, HDI = 0.12–1.29), confirming statistical relevance.

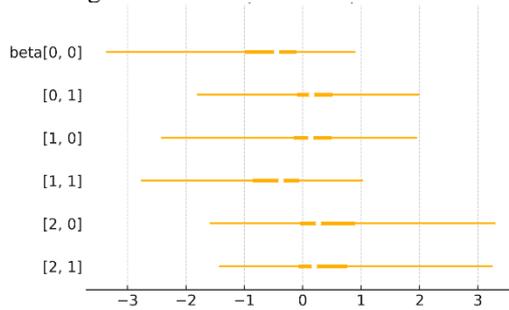
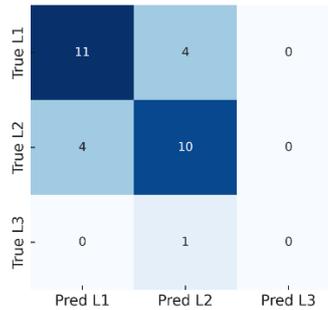

**Fig. 7.** Posterior 95% HID for $\beta$ Coefficients

**Fig. 8.** Confusion Matrix

Practically, the model indicates that crossing lamellae and additional wet cycles act synergistically to lower reuse level. Because the other three candidate features collapse toward zero effect, the minimal on-site predictor set is reduced to just two binary entries which are the cycles of the member exposed to and bending lamella run long-grain or cross-grain, which eliminating the need to measure continuous properties on site is advantageous for rapid triage.

### 3.5 Classification accuracy and uncertainty

Using posterior-mean logits, the categorical predictions yield the confusion matrix shown in **Fig. 8**. Overall accuracy ($(TP + TN)/N$) is 67 %; misclassifications are almost exclusively off-by-one-level—only one Set 2 sample (4 % of that group) is erroneously labelled Level 1. The multiclass Brier score, which penalises probabilistic miscalibration, equals 0.432 (perfect score = 0, worst = 2 for three classes).

## 4 Discussion

The single wet–dry exposure (Set 1) reduced the weighted residual-performance metric $R$ by 7 % on average, yet 68 % of specimens still exceeded the 0.90 Level-1 threshold. A second exposure (Set 2) compounded the deterioration: mean $R$ fell to 0.78 and the share of Level-1 members halved to 33 %. The posterior coefficients confirm that degradation is magnified in cross-laminated orientations, where swelling of transverse lamellae concentrates shear stresses during drying. Consequently, Set 2–cross boards exhibit only a 21 % probability of direct reuse versus 48 % for Set 1–cross and 74 % for Set 1–long members. The Bayesian model reduces quality-assurance inputs to two binary variables—wet-cycle count and grain orientation—yet still quantifies uncertainty for every component.A simple decision rule (**Fig. 9**) leveraging $P(Level\ 1)$ allows rapid triage: $P \geq 0.70$ : redeploy as Level 1 without further tests; $0.40 \leq P < 0.70$ : perform a non-destructive check (e.g., ultrasound); $P < 0.40$ : downgrade or reject.

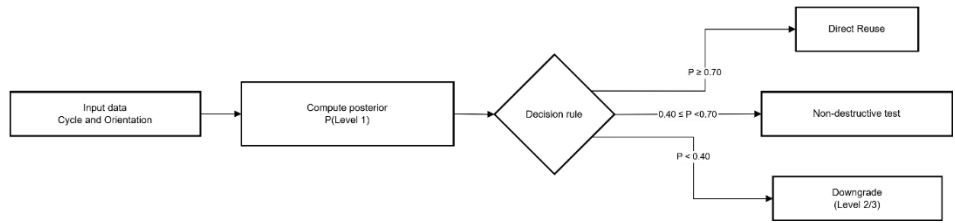

**Fig. 9.** Cycle Prediction Flow

## 5 Conclusions

This study offers the first probabilistic framework for classifying *moisture-exposed engineered timber* within the MMC reuse hierarchy. Flexural tests on 72 spruce CLT specimens showed that a single soak–dry cycle lowers the combined stiffness-and-strength metric $R$ by 7 %, while two cycles cause a 22 % loss. Using these data we trained a Bayesian multinomial-logistic model regularised by horseshoe priors. Remarkably, only two binary inputs, *wet-cycle count* and *grain orientation,* were needed to reproduce the observed level distribution. Replacing the current moisture-limit rule with the proposed probability-based classifier would approximately double the volume of salvaged timber that can be reused without structural downgrading, yet do so under explicit, sample-specific uncertainty. However, this work is subject to principal constraints like limited specimens, single laminate type, destructive testing and limited moisture protocol. Further research will be conducted on multi-cycle exposure to reveal asymptotic degradation trends and permit calibration of a fatigue term in the probabilistic model.